\documentclass{llncs}
\usepackage{enumerate}
\usepackage{amssymb}
\usepackage{url}
\usepackage{tabularx}
\usepackage{amsmath}
\usepackage{commath}
\usepackage{hyperref}
\usepackage{listings}
\usepackage{caption}
\usepackage{subcaption}
\usepackage{colortbl}
\usepackage{multicol}
\usepackage{xcolor}
\usepackage{dblfloatfix}
\usepackage{paralist}
\usepackage{csquotes}
\usepackage{balance}
\usepackage{amsmath}
\usepackage{amssymb}
\usepackage{booktabs}
\usepackage{multirow}
\usepackage[noend]{algorithmic}
\usepackage{algorithm}

\algsetup{indent=2em,linenodelimiter=.,linenosize=\small} % indentation

\newcommand{\hide}[1]{}

\usepackage{graphicx}
\usepackage{booktabs}
\usepackage[utf8]{inputenc}
\usepackage{hyperref}
\usepackage{makecell}
\usepackage[skip=10pt]{caption}
\usepackage{listings}

\begin{document}

\title{RAPTOR: Ransomware Attack PredicTOR}
\titlerunning{RAPTOR}  % abbreviated title (for running head)
\author{Florian Quinkert\inst{1} \and Thorsten Holz\inst{1} \and KSM Tozammel Hossain\inst{2} \and Emilio Ferrara\inst{2} \and Kristina Lerman\inst{2}}
\authorrunning{Florian Quinkert et al.} % abbreviated author list (for running head)
\tocauthor{Florian Quinkert, Thorsten Holz, KSM Tozammel Hossain, Emilio Ferrara, Kristina Lerman}
\institute{Ruhr-Universität Bochum, Germany, \email{firstname.lastname@rub.de}
\and
USC Information Sciences Institute, Los Angeles, USA,\\
\email{tozammel@isi.edu}, \email{emiliofe@usc.edu}, \email{lerman@isi.edu}}

\maketitle              % typeset the title of the contribution

\begin{abstract}
%To paraphrase an old saying, ``foreknowledge is the best defense'': this has become increasingly the case in the fight against ransomware. In this paper, we present RAPTOR, a promising line of defense based on fingerprinting and forecasting the operations leading to ransomware attacks. We monitor newly registered domains and use a machine learning model to detect suspicious ones. Building upon timeseries techniques, we learn models of historical ransomware attacks, and use them to predict future cyber threats, including malicious host registrations and sequences of ransomware attacks.
%We illustrate the effectiveness of our framework by forecasting all stages of attacks based on Cerber, a popular ransomware: by looking at the zone files of the top-level domain \texttt{.top} starting from 2016/08/30, we predicted 2126 Cerber domain candidates. 378 of those domains were later observed in blacklists. We finally demonstrate how our timeseries forecasting models can accurately predict the volume of malicious hosts registrations, as well as anticipating the number of daily Cerber attacks, during a two-months period at the end of 2016.

Ransomware, a type of malicious software that encrypts a victim's files and only releases the cryptographic key once a ransom is paid, has emerged as a potentially devastating class of cybercrimes in the past few years. 
In this paper, we present RAPTOR, a promising line of defense against ransomware attacks. 
RAPTOR fingerprints attackers' operations %and  uses machine learning 
to \emph{forecast} ransomware activity. More specifically, our method learns features of malicious domains by looking at examples of domains involved in known ransomware attacks, and then monitors newly registered domains to identify potentially malicious ones. In addition, RAPTOR uses time series forecasting techniques to learn models of historical ransomware activity and then leverages malicious domain registrations as an external signal to forecast future ransomware activity. 
We illustrate RAPTOR's effectiveness by forecasting all activity stages of Cerber, a popular ransomware family. By monitoring zone files of the top-level domain \texttt{.top} starting from August 30, 2016 through May 31, 2017, RAPTOR predicted 2,126 newly registered domains to be potential Cerber domains. Of these, 378  later actually appeared in blacklists. Our empirical evaluation results show that using predicted domain registrations helped improve forecasts of future Cerber activity. Most importantly, our approach demonstrates the value of fusing different signals in forecasting applications in the cyber domain.

\end{abstract}
\section{Introduction}
\hide{KL: Points to make:
\begin{itemize}
\item Describe the lifecycle of an attack: variant deployment, registration of host, detection of infection
\begin{itemize} 
\item 2 phases of ransomware (infection \& payment)
\item  infection: malspam \& exploit kit (3 parts: injected script, exploit kit, ransomware)
\item  payment: C\&C requires communication with host, newly registered or existing compromised host

\end{itemize}
\item Fingerprints of processes allow us to learn patterns
\begin{itemize}
\item Registration: host name name, registrant features, temporal patterns of registration
\item Detection: temporal patterns
\end{itemize}
\item Case study: Cerber ransomware
\item Generalize to other malware types (other ransomware, phishing, etc.)
\end{itemize}
}
%Ransomware has been around since the mid 2000s. However, it got widely adopted only a few years ago. 
Ransomware has emerged as a potentially devastating class of cybercrimes. In a ransomware attack, an adversary tricks victims into downloading malicious software that blocks access to their computer systems until a sum of money, or a similar ransom, is paid.
Attacks on businesses have grown dramatically over the past few years. In the first quarter of 2017, Kaspersky blocked more than 240.000 ransomware infections on unique users' computers and detected 11 new ransomware families and more than 55.000 variants. The most prevalent ransomware family was Cerber, which attacked more than 18\% of the victims~\cite{kaspersky-ransomware-q1-2017}. In May 2017, a vulnerability in Microsoft Windows operating systems exposed hundreds of thousands of computer systems worldwide to the WannaCry ransomware~\cite{kryptos_17_05_29}, causing widespread disruption of central infrastructure and services, including hospitals, police departments, etc.
%
%Such proliferation poses significant challenges for defending against ransomware, for example, through antivirus signatures and similar reactive mechanisms.

%In this paper, we present RAPTOR (Ransomware Attack PredicTOR), a novel solution to the challenge of defending against ransomware attacks by forecasting them. Our approach relies on modeling attacker behavior within a life-cycle of an attack and using the learned patterns to predict new attacks or their properties. The intuition driving our approach is that attacker activities need to be scalable to be productive. To achieve scalability, attackers often rely on automated processes, the traces of which are visible in domain registration logs, blacklists, etc. By analyzing their traces, we can fingerprint these  processes, which will in turn, enable us to anticipate attacker behavior.  

%As an illustration, consider 
The life-cycle of a ransomware attack is divided into two phases: an infection phase and a payment phase. In the infection phase, the %ransomware has to get access to
attacker penetrates 
the victim's computer,
%For that purpose, two different infection mechanisms are possible,
which is typically accomplished through 
malicious e-mails or via Websites infected with exploit kits (so called \emph{drive-by download attacks}). In the case of phishing e-mails, the attacker sends an e-mail to the victim with a malicious attachment. When the victim opens the attachment, the computer gets infected, and the ransomware encrypts the victim's hard drive with a strong encryption algorithm. In case of drive-by download attacks, an adversary injects a small piece of code into a legitimate Website so that visitors are directed to the landing page of an exploit kit. In most cases, these injections are performed by scripts that crawl and search for vulnerabilities in large amounts of Websites. The exploit kit checks the victim's browser for vulnerabilities it can target and delivers the ransomware as payload, which then encrypts the victim's hard drive. 

After a successful infection, the ransomware displays a ransom note and demands a payment from the victim for %the victim to receive 
the decryption key. %The ransomware has 
In order to keep track of incoming payments, the ransom note contains one or more URLs to payment Websites. The URLs typically contain a static part %which is equal for a group of users 
and a variable part, which includes %in most cases 
a unique identifier. As it is very important to the attacker to ensure anonymity of payments, the URLs either refer to a hidden service in the Tor network or to a Website in the clearnet which runs a service like Tor2Web, %In the former case, users have to use Tor browser to access the website. Tor2Web 
which enables users to visit Websites in the Tor network without a Tor browser (although without providing the strong anonymity features typical of Tor). 

In most cases, the ransomware developer changes the static part of the URL regularly, which requires them to frequently register new domains. However, since the registration of new domains is a tedious task, most ransomware developers either rely on scripts for the registration process or use identifiable patterns during the registration that we can leverage in the analysis process.

In 2016, Hao \textit{et al.} proposed a system called \textsc{Predator} to predict whether a domain will be used maliciously, based on the features %inferred 
extracted at registration time, such as %structure 
patterns of the registered domains~\cite{hao2016predator}.  
Remarkably, 70\% of the domains predicted by \textsc{Predator} were later observed in domain blacklists. However, \textsc{Predator} can only distinguish between malicious and benign domains and does not provide information about the purpose of a particular domain, e.g., whether it is used by a certain ransomware campaign.

In addition, patterns in the time series of ransomware activity may be useful for predicting future ransomware activity. These patterns may arise when developers continue to utilize a  specific ransomware variant until antivirus defenses catch up, requiring them to develop a new variant. In addition, the events at different stages of ransomware life cycle are linked: attackers register new domains when they expect them to be needed, i.e., when attacks are taking place. Temporal correlations in ransomware activity, and between different stages of ransomware activity, have predictive value. The goal of time series prediction is to analyze historical  activity to identify patterns, which can then be used to infer future activity.

%\note{Our contributions.}
Inspired by these observations, we present \textsc{RAPTOR}---Ransom\-ware Attack PredicTOR---a novel framework to predict ransomware activity. \textsc{RAPTOR} consists of two closely interconnected components. The first component identifies potential ransomware domains at time-of-registration using the observed patterns, while the second component uses these domains as an external signal in time series forecasting to predict future ransomware activity.

%and domain usage new ransomware attacks by analyzing existing attacks to fingerprint attacker behaviors. 

In the first part of this paper, we describe how to learn features of malicious domains by leveraging information in registration logs, enriched with WHOIS information. We show that these features can effectively identify new malicious domains at the time of registration. %We also show how to use time series analysis to predict the number of new malicious domains. 
Afterwards, we describe a method for time series forecasting that uses predicted malicious domains as an external signal to improve predictions of ransomware activity. 
We validate our approach by applying it to the popular Cerber ransomware. % In a real-world setting, over the period from August 2016 to May 2017, RAPTOR predicted 2126 newly registered domains to be potential Cerber domains. Of these, 378  later appeared in blacklists.  

In summary, we make the following contributions:
\begin{itemize}
\item \emph{Malicious domain prediction}: We demonstrate how ransom\-ware campaigns can be characterized to extract features that are predictive of new malicious domains. More specifically, we focus on information available via the Domain Name System (DNS) and WHOIS to identify likely domains that will become active in the future.
\item \emph{Time series forecasting}: We study how external signals can be added to time series forecasting to improve predictions of future ransomware activity.  
\item \emph{Predicting ransomware domains% associated with ransomware attacks
}: We evaluate \textsc{RAPTOR} with a detailed case study of the ransomware family Cerber. We show that both techniques can predict future events, and we were able to successfully predict 378 domains that eventually showed up in malware blacklists.
\end{itemize}

The paper is organized as follows. We review existing literature in Section~\ref{related}. Next, we describe \textsc{RAPTOR}'s components in Section~\ref{approach}. Afterwards, we evaluate \textsc{RAPTOR} with reference to the ransomware family Cerber in Section~\ref{case_study_cerber} and conclude with a discussion of the findings and future directions in Section~\ref{conclusion}. 
\section{Related Work}
\label{related}
Our work is based on previous research on domain life-cycle and time series prediction and we now briefly review related work to discuss how it relates to our approach. Additionally, we review existing literature about ransomware detection. \par

\textit{Domain life-cycle} Recent research analyzed different misbehaviors related to domain registrations. Coull \textit{et al.} described malicious domain registration techniques, such as \textit{domain tasting} and \textit{domain front running}~\cite{coull2010understanding}. Alrwais \textit{et al.} examined domain parking services and detected \textit{click fraud}, \textit{traffic spam} and \textit{traffic stealing} from the parked domains~\cite{alrwais2014understanding}. Szurdi \textit{et al.} analyzed \textit{typosquatting} in the top-level domain \textit{com} and estimated that about 20\% of the registered domains were typosquatted domains~\cite{szurdi2014long}. Agten \textit{et al.} analyzed \textit{typosquatting} of the 500 most popular domains over seven months in 2013 and revealed that about 95\% of the domains were targeted by \textit{typosquatting}~\cite{agten2015seven}.

Besides reactively analyzing malicious domain activities, proactively determining domains registered for malicious activities received  more  attention in recent years. Felegyhazi \textit{et al.} inferred features from a known bad domain and found 3.5 to 15 new domains, of which about 74\% ultimately occurred in blacklists~\cite{felegyhazi2010potential}. Hao \textit{et al.} analyzed the domain registration process of domains used in spamming campaigns and inferred features to determine such domains already at time of registration~\cite{hao2013understanding}. Later on, Hao \textit{et al.} presented \textsc{Predator}, a system to predict at time of registration whether domains will be used for malicious purposes~\cite{hao2016predator}. We extend this line of work and demonstrate how it can be tailored to predict ransomware activity. \par

\textit{Time series prediction}
Developing a precise model for the dynamic behavior of time series is a challenging problem and an essential one for the success of forecasting methods. Researchers have extensively studied and used time-series analysis in many domains, such as finance~\cite{lendasse2000non}, epidemiology~\cite{chakraborty2014forecasting}, geophysics~\cite{shumway2010time}, and sociology~\cite{box2014time}. A popular strategy for analyzing time series data is using  classical autoregressive models such as AR, ARMA, ARIMA, and ARIMAX~\cite{box2015time,shumway2010time,prado2010time}. Autoregressive models are widely used in intrusion detection, detecting DoS attacks, and network monitoring~\cite{viinikka2009processing}. These models assume that the underlying data-generating process is linear, \textit{i.e.}, the value at a time point is a linear combination of the past values. However, real-world time series exhibit volatility and nonlinearity. A way to deal with the problem of volatility is to employ ARCH and GARCH, which are extensions of classical autoregressive models~\cite{douc2014nonlinear}. To address the problem of nonlinearity, we can exploit state space models, such as hidden Markov models~\cite{rabiner1986introduction} and dynamic linear models~\cite{shumway2010time}. A hidden Markov model (HMM) assumes that the dynamics of a system at a time point are generated by one of the possible hidden states (unobserved regimes) evolving according to a Markov chain over time~\cite{douc2014nonlinear}. HMM models are exploited to infer interpretable threat trends and detecting attacks from time series data~\cite{kim2007cyber,ye2004robustness}. \par

\textit{Ransomware detection}
Ransomware is one of the most prevalent threats. Therefore, researchers recently proposed different mechanisms to detect ransomware. Kharaz \textit{et al.} proposed UNVEIL, a system that detects ransomware by keeping track of typical ransomware behavior like massive encryption of files~\cite{kharraz2016unveil}. They were able to detect more than 13,000 malware samples from different families. Andronio \textit{et al.} presented HELDROID, which searches for necessary ransomware components in mobile applications~\cite{andronio2015heldroid}. Continella \textit{et al.} described ShieldFS, a filesystem add-on that detects and rolls back malicious changes~\cite{Continella:2016:SSR:2991079.2991110}. Kolodenker \textit{et al.} introduced PayBreak, a system that detects the usage of symmetric cryptography and stores the used session keys to provide the user access to the encrypted data~\cite{Kolodenker:2017:PDA:3052973.3053035}.

In contrast to the presented related work, our approach focuses not on the detection of ransomware infections on a host system but on the forecasting of ransomware campaigns.

% We will broadly use timeseries prediction models of these various families throughout the rest of the paper.
\section{RAPTOR}
\label{approach}
For their attacks to scale, ransomware developers rely on scripts to automate (parts of) the ransomware life-cycle. Our goal is to characterize attackers by observing the traces their scripts and other malicious behaviors leave. The intuition is that these traces enable us to learn features that are predictive of new attacks. Specifically, we want to learn features which will allow us to predict---at the time of registration---whether a domain will eventually be used in a ransomware attack. We also want to predict how many such domains will be registered. In the following, we describe our approach, called \textsc{RAPTOR}, to address these challenges.

% {\color{red}{{Todo: use domains to predict attacks}}}

\subsection{Ransomware Domain Prediction}
We refer to domains right after registration as \textit{inactive} and to those actually being used by ransomware developers as \textit{active}. According to RFC 1034~\cite{rfc1034}, a zone file consists of resource records which contain the domain name, type, class, time to live, and RDATA fields. We download zone files on a daily basis from ICANN's Centralized Zone Data Service (CZDS)~\cite{czds} and compare the current zone file to the previous day's zone file to identify newly added resource records. 

Additionally, we collect active domains, which were used by a particular ransomware from blacklists, such as \textit{abuse.ch}'s ransomware tracker~\cite{abusech}. We use these domains together with known benign domains to train a supervised machine learning based classifier in order to predict which of the newly registered domains will be used by the ransomware.

\begin{figure}[t]
  \centering
  \includegraphics[width=0.6\textwidth]{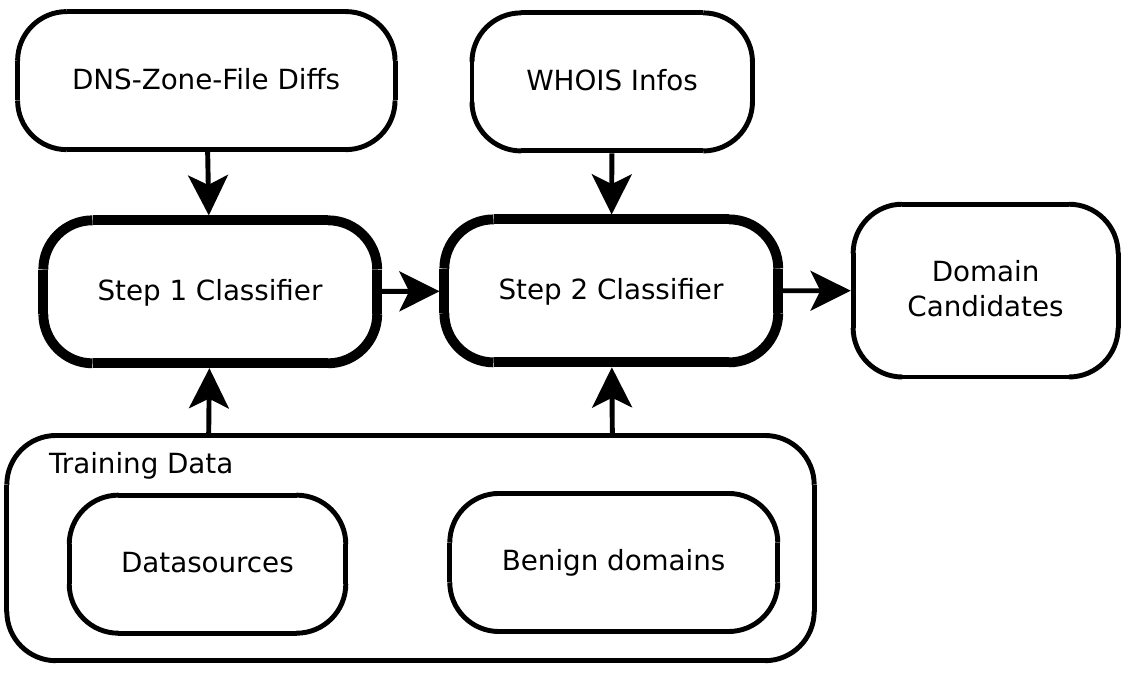}
  \caption{High-level overview of \textsc{RAPTOR}'s  classifiers}
  \label{raptor_high_level_overview}
\end{figure}

Figure~\ref{raptor_high_level_overview} shows a high-level overview of RAPTOR's domain classifiers. In particular, we use two classifiers to predict which domains will be used by the ransomware First, we use a classifier which filters newly registered domains from the zone file diffs based on the domain structure and the nameservers, i.e., information obtained from the zone file itself (\textit{Step 1 Classifier}). Afterwards, we collect WHOIS information about the remaining domains and use a second classifier to filter based on this information (\textit{Step 2 Classifier}). 

We use the same sets of known malicious and known benign domains as training data for both classifiers. However, we use different features.
The WHOIS service provides information about a domain, such as the registrant, the admin, and/or the time of registration~\cite{rfc3912}. Collecting large amounts of WHOIS information is a time-consuming task, since the WHOIS service stops serving responses after a small number of requests is sent by the same IP address in a short period of time. Additionally, the response's structure is different for each top-level domain so that multiple parsers are necessary. Commercial vendors offer parsed WHOIS information for sale. However, in most cases the number of requests per day/month is rather small. %, especially for the requested price. 
We obtain our WHOIS information from a service called whoisxmlapi~\cite{whoisxmlapi}. % which granted us an academic discounted license.
Therefore, we use the \textit{Step 1 Classifier} to reduce the number of 
potential malicious domains, i.e., the number of necessary WHOIS requests, and the \textit{Step 2 Classifier} to predict which domains are  
likely to be used maliciously by 
the particular ransomware. We refer to these domains as \textit{domain candidates}.

The \textit{Step 1 Classifer} and \textit{Step 2 Classifier} use features shown in Table~\ref{features}. Since ransomware developers use scripts to register a large number of domains, the features aim at finding patterns in the domain structure and WHOIS information of the training data to apply them on the newly registered domains. Features \#1 - \#7 focus on patterns that are directly inferrable from the domain name or WHOIS information. Features \#7, \#10 and \#11 use common reuse of information during the registration process, whereas features \#8 and \#9 use common features of malicious domain registrations, such as a rather short registration time period.

\begin{table}[]
\centering
\caption{Features used in \textit{Step 1 / Step 2 Classifier}}
\label{features}
\begin{tabular}{@{}ll|ll@{}}
\toprule
\multicolumn{2}{l}{\textit{Step 1 Classifier} features} & \multicolumn{2}{l}{\textit{Step 2 Classifier} features} \\ \midrule
\# & Feature description & \# & Feature description    \\
1  & Length of domain name               & 7 & Registrant's organ. is part of registrant's name\\
2  & At least one letter in domain name  & 8 & Number of days a domain is registered for       \\
3  & Domain name consists only of digits & 9 & Weekday of registration                         \\
4  & Number of distinct characters       & 10 & Registrant's fax equals registrant's telephone \\
5  & Hyphen in domain name               & 11 & Registrant, admin and tech are equal           \\
6  & Domain name starts with number                         \\ \bottomrule
%7  & Nameserver is known from training data                 \\
%\midrule
%\multicolumn{2}{l}{\textit{Step 2 Classifier} features}     \\
%8  & Registrar is known from training data                  \\
%9  & Registrant's name is known from training data          \\
%10 & Registrant's email is known from training data         \\
%11 & Registrant's postal code is known from training data   \\
%12 & Registrant's street is known from training data        \\
%13 & Registrant's country is known from training data       \\
%14 & Registrant's state is known from training data         \\
%15 & Registrant's city is known from training data          \\
%16 & Registrant's fax is known from training data           \\
%17 & Registrant's telephone is known from training data     \\
%7 & Registrant's organization is part of registrant's name  \\
%8 & Number of days a domain is registered for               \\
%9 & Weekday of registration                                 \\
%10 & Registrant's fax equals registrant's telephone         \\
%11 & Datasets for registrant, admin and tech are equal      \\ \bottomrule
\end{tabular}
\end{table}

\hide{
outline:
formally define time series and the problem.
models
--HMM
--arima
--zip

base-rate models
--current count as a forecast
--rolling average
}

\subsection{Time Series Prediction}
\label{sec:ts-models}

The intuition behind time series prediction is that when events are correlated in time, then given a sequence of events, 
%it is possible to learn a model characterizing the process that generated these events and use the learned model to predict new events. 
one can learn patterns of past events that are useful for predicting future events. 
For example, consider that a ransomware attack has already occurred. An existing attack increases the likelihood of another attack, since ransomware developers reuse the tools and kits to launch repeated attacks or attacks on new targets. Time series prediction techniques use historical data about events to learn a model of the process that produced these events. The model can, in turn, be used to predict new events. Below we describe how we apply two models widely used in time series prediction---the hidden Markov model and the autoregressive model---to address the challenge of modeling malicious behavior.

\subsubsection{Prediction with the Hidden Markov Model}
\label{sec:hmm-model}
\begin{figure}[ht]
  \centering
  \includegraphics[scale=0.5]{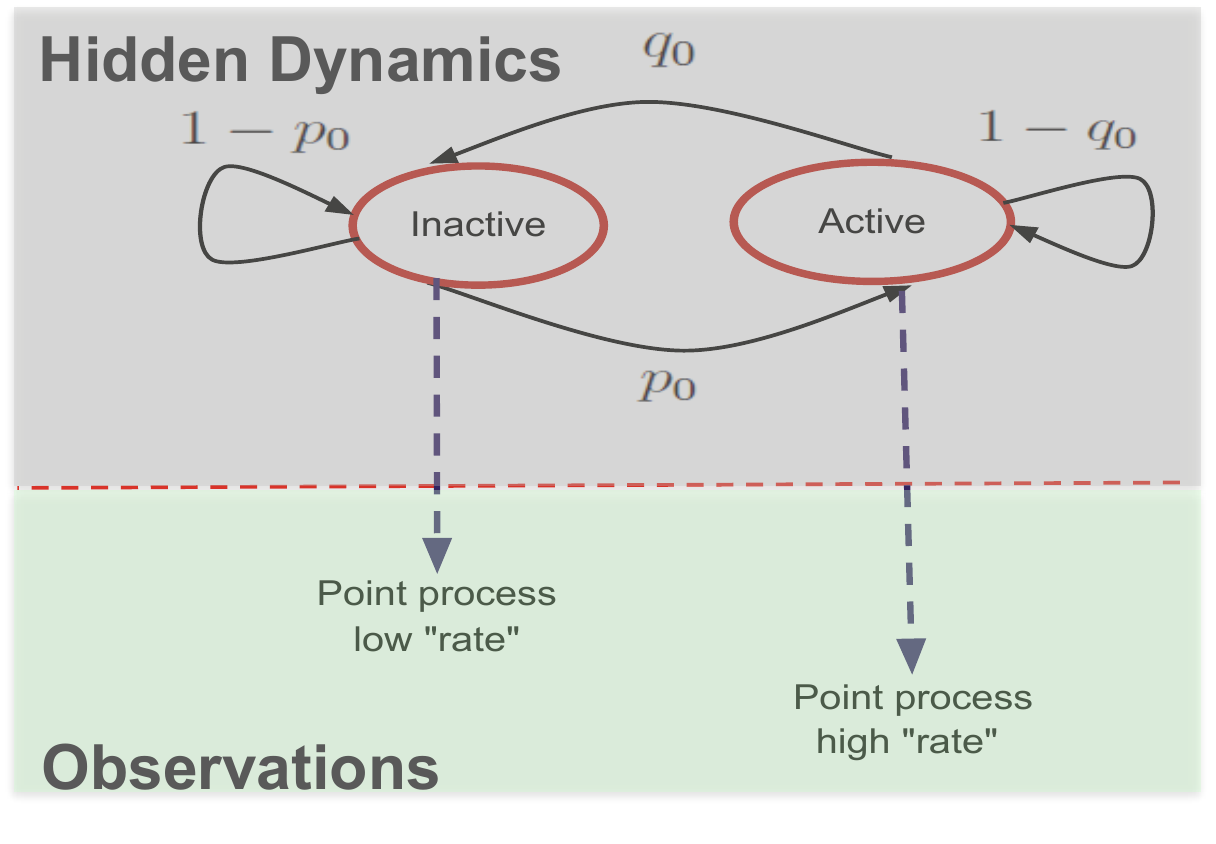}
  \caption{Hidden Markov model for ransomware domain registration.}
  \label{fig:hmm}
\end{figure}

We propose to use a hidden Markov model (HMM) to predict malicious domains that will be involved in attacks.  HMMs have been successfully used for example to model and predict attacks of terrorist groups \cite{raghavan2013hidden}.
In our context, the key idea of HMM is that the current number of events (e.g., detected domains) depends on past history of events through $K$ dominant hidden states, which represent different operational phases of the registration process. For example, the hidden states of a two-state HMM correspond to `low-activity' and `high-activity' processes, as shown in Figure~\ref{fig:hmm}. The process transitions probabilistically between a low-activity and high-activity states. While in a particular state, the process outputs some events according to a statistical distribution but with a state-dependent rate.

% An HMM is described by a set of hidden states, transition
% probabilities between hidden states, and observed probabilities condition on the hidden
% state.
Let $\mathcal{Y}=(y_1,y_2,\ldots,y_T)$ be the observed sequence of events, e.g., the daily number of ransomware domain registrations or attacks,  and
$\mathcal{Z}=(z_1,z_2,\ldots,z_T)$ be the underlying states of the process giving rise to the events $\mathcal{Y}$. Here $T$ denotes the length of the time series, i.e., the sequence of events.  
An HMM is described by a set of hidden states ($\mathcal{S} = \{S_1,
S_2,\ldots,S_N\}$), transition probabilities between the states ($\eta_{ij}
= P(z_t = S_j|z_{t-1}= S_i)$), initial probabilities of 
the states ($\pi_i = P(z_1 = S_i)$), and the emission probabilities of events conditioned on the hidden state ($\phi_{i}(k) = P(y_i=k|z_t=S_i)$).  The hidden states $\mathcal{Z}$ are discrete-valued random variables.  A transition between the states is Markovian, i.e., the future state is conditionally independent of the past states given the current state. In our problem setting, we consider the emission probabilities of events to be a continuous value from one of  four possible distributions: Poisson, Gaussian, geometric, and Hurdle geometric. The generative process for the model is shown in Algorithm~\ref{alg:genprocess}. \par

\begin{algorithm}[!t]
  \caption{Generator($\eta,\pi$) for HMM}
  \label{alg:genprocess}
  \begin{algorithmic}[1]
    \REQUIRE {A set of parameters.}
    \ENSURE {Number of domain registrations.}
    %\medskip
    \STATE{Choose the initial state $z_1 \sim \text{Mult}(\pi)$}
    \STATE{Draw each row of $\eta_i$ using $\text{Dir}(\alpha)$}
        \COMMENT{Transition matrix for a user-defined $\alpha$}
    \STATE {Choose the emission probability distribution $\phi \in 
        \{\text{Poisson, Gaussian, Geometric, Hurdle Geometric}\}$ }
    \FOR{each time $1 \le t \le T$}
        \IF {not the 1st day} 
            \STATE{$z_t \leftarrow \text{Mult}(\eta_{z_{t-1}})$}
        \ENDIF
        \STATE{Draw $y_t \sim \phi_{z_t}$}
    \ENDFOR
  \end{algorithmic}
\end{algorithm}

\textit{Estimating HMM Parameters} The unknown parameters of the proposed HMM model are $H = \{\pi,\eta,\phi\}$. No analytical solution exists for this model that maximizes the
probability of the observed sequence (i.e., likelihood)~\cite{rabiner1989tutorial}. Hence, we applied an Expectation Maximization (EM) based algorithm (a.k.a Baum-Welch reestimation) to
estimate the parameters of the model. \par

\textit{Predicting with HMM}
To predict the number of new events (i.e., ransomware domain registrations), 
we adopt a sliding window approach. We learn our model with data determined by a user-defined time window (e.g., 50 days) and forecast the expected number of events for the next day, to the last day of the time window. The expected number of events at time $t$ given $z_{t-1}$ is  
\begin{align}
\bar{y}_t = \sum_j^{N} \eta_{z_{t-1}j} * \text{E}[S_j],\nonumber
\end{align}
where $\text{E}[S_j]$ is the expected number of events at state $S_j$.

\subsubsection{Prediction with the ARIMA Model}
\label{sec:arima}
We also present sliding-window based autoregressive models (ARIMA) for forecasting events, including \hide{domain registrations and} reported domains in ransomware attacks. ARIMA stands for autoregressive integrated moving average. The key idea is that the number of current events ($y_t$) depends on the past counts and forecast errors. Formally, ARIMA($p$,$d$,$q$) defines an autoregressive model with $p$ autoregressive lags,
$d$ difference operations, and $q$ moving average lags (see~\cite{shumway2010time}). Given the observed series of events $\mathcal{Y}=(y_1,y_2,\ldots,y_T)$, ARIMA($p$,$d$,$q$) applies 
$d$ ($\ge 0$) difference operations to transform $\mathcal{Y}$ to a stationary series
($\mathcal{Y}^{\prime}$). Then the predicted value $y^\prime_t$ at time point $t$ can be expressed in terms of past observed values and forecasting errors which is as follows: 
\begin{align}
    y^\prime_t = c + \sum_{i=1}^p \alpha_i y^\prime_{t-i} + \sum_{j=1}^q \beta_j e_{t-j} + e_t\label{eq:arima}
\end{align}
Here $c$ is a constant, $\alpha_i$ is the autoregressive (AR) coefficient at lag $i$, $\beta_j$ is the moving average (MA)
coefficient at lag $j$, $e_{t-j} = y^\prime_{t-j} - \hat{y}^\prime_{t-j}$ is the forecast error at lag $j$, and $e_t$ is assumed to be the white noise ($e_t\sim \mathcal{N}(0,\sigma^2)$). \hide{The AR model is essentially an ARIMA model without moving average terms}. 

We use maximum likelihood estimation for learning the parameters; more specifically, parameters are optimized with LBFGS method~\cite{seabold2010statsmodels}. These models assume that $(p, d, q)$ are known and the series is weakly stationary. To select the values for $(p, d, q)$ we employ grid search over the values of $(p, d, q)$ and select the one with minimum AIC score. A key motivation for using ARIMA is that the model requires limited history for forecasting.

\subsubsection{Prediction with the ARIMAX Model}
\label{sec:arimax}

ARIMAX (Autoregressive Integrated Moving Average with Exogenous variables) is an autoregressive model that leverages (optional) external signals. In this model, the observation at a particular time point depends on immediate past observations, past forecast errors, and external variables. Like ARIMA, ARIMAX is defined with three order terms: a) number of autoregressive order ($p$), b) number of difference operations used to make the series stationary ($d$), and c) number of moving average terms ($q$) (see~\cite{shumway2010time}). Given the observed series of events $\mathcal{Y}=(y_1,y_2,\ldots,y_T)$ and optional $K$ external features\\
$\mathcal{X} = (\langle x_{11},x_{21},\ldots,x_{K1}\rangle,\allowbreak \langle x_{12},x_{22},\ldots,x_{K2}\rangle,\ldots, \langle x_{1T},x_{2T},\ldots,x_{KT}\rangle)$,\allowbreak
the model is defined as follows:
\begin{align}
    y^\prime_t = c + \sum_{i=1}^p \alpha_i y^\prime_{t-i} + \sum_{j=1}^q \beta_j e_{t-j} + \sum_{k=1}^K \gamma_k x_{kt} + e_t\label{eq:arimax}
\end{align}
Here $\mathcal{Y}^{\prime}$ is the stationary series after $d$ difference operations, $c$ is a constant, $\alpha_i$ is the autoregressive (AR) coefficient at lag $i$, $\beta_j$ is the moving average (MA)
coefficient at lag $j$, $e_{t-j} = x^\prime_{t-j} - \hat{x}^\prime_{t-j}$ is the forecast error at lag $j$, $\gamma_k$ is the coefficient for feature $x_k$ and $e_t$ is assumed to be the white noise. 

We apply maximum likelihood and sliding window approaches for estimating model  parameters and evaluation, respectively . Our motivation to use ARIMAX for predicting domains involved in ransomware activity is the observation of correlation between domain registration and detection of domains used by ransomware at various time lags. In addition to the requirement of limited history for prediction, this model harnesses external signals and is robust to the absence of historical data.

%We trained the model with three months of data and test over an entire month. While sliding over the test period, we do not retrain the model. 
%Forecasting based on historical data may not lead to a better forecasting accuracy. 

\subsubsection{Base rate model}
\label{sec:base-rate}
We compare our proposed methods discussed in Sec.~\ref{sec:hmm-model} and Sec.~\ref{sec:arima} against a base rate model, which is a rolling average model:
\textit{Rolling Average} predicts that the number of future events will be the average number of past events over a time  window $W$.
%\begin{align}
$    x^\prime_t = \frac{1}{W}\sum_{i=1}^W x_{t-i}.$ 

%a) current count and b) rolling average. The current count method predicts the future events with the most up-to-date count and the rolling average method predict the future events with an average number of events over a window of past events. Formally, these methods can be defined as follows:

%\begin{itemize}
% \item \textbf{Current Count}  predicts that the number of future events will be the same as the number of current events.
% %\begin{align}
% $    x^\prime_t = x_{t-1}.$ 
% %\end{align}
%\item 
%\end{align}
%\end{itemize}

\subsubsection{Evaluation of time series models}\label{sec:ts-metrics}
We use three error measures for quantitative evaluation of our time series models: a) mean absolute error (MAE), b) root mean squared error (RMSE), and c) mean absolute scaled error (MASE)~\cite{hyndman2006another}. These measures are defined as follows in terms of forecasting error, $e_t = y_t - y\prime_t$, at time point $t$, where $y_t$ and $y\prime_t$ are the true and predicted values, respectively.

%\begin{itemize}
% \item $\textbf{MAE} = \frac{1}{T}\sum_{t=1}^{T}\abs{e_t}$ 
% \item $\textbf{RMSE} = \sqrt[]{\frac{1}{T}\sum_{t=1}^{T}\abs{e_t}^2$} 
% \item $\textbf{MASE} = \frac{\frac{1}{T} \sum_{t=1}^T|e_t|}{
% 		\frac{1}{T-1} \sum_{t=2}^T|y_t - y_{t-1}|
% 		}$
% MAE
%\item \textit{Mean Absolute Error}
%\item[]
\begin{align*}
	\text{MAE} = \frac{1}{T} \sum_{t=1}^T|e_t| \qquad
    \text{RMSE} = \sqrt{\frac{1}{T} \sum_{t=1}^T|e_t|^2} \qquad
  	\text{MASE} = \frac{\frac{1}{T} \sum_{t=1}^T|e_t|}{
                        \frac{1}{T-1} \sum_{t=2}^T|y_t - y_{t-1}|
		          }
\end{align*}

\section{Evaluation}%{Case study: Cerber}
\label{case_study_cerber}
We have implemented a prototype of the approach discussed in the previous section in a tool called \textsc{RAPTOR}. In the following, we evaluate our approach on the ransomware Cerber and conduct both a real-world case study incorporating zone files collected between August 2016 and May 2017, and a cross-validation with the domains already known to be used by Cerber. We evaluate both scenarios with reference to precision (number of true positives divided by sum of true and false positives), recall (true positives divided by expected true positives) and f1-score (the geometric mean of precision and recall). %two times product of precision and recall divided by sum of precision and recall). 
We also evaluate the time series models on the task of predicting the number of domains involved in Cerber attacks using the past time series of such domains, along with time series of malicious domains predicted by our method. 

\subsection{Ransomware Cerber}\label{sec:data-source}
%We start by providing general information about Cerber. %Afterwards, we analyze our dataset of known Cerber domains and introduce the inferred features for the feature learning. Eventually, we present and evaluate our prediction results. 
Cerber was first detected at the beginning of March 2016~\cite{cerber-first-occurrence}. It infects victims via both malicious spam e-mails~\cite{malspam-example} and exploit kit-infected websites~\cite{exploit-kit-example}. Cerber encrypts the data on a victim's computer with AES encryption and asks for a ransom to get the decryption key. The ransom note contains between one and four different URLs for the ransom payment. The URLs link either to TOR Onion Services, accessible only with the Tor browser, or to websites running the Tor2web service~\cite{tor2web}, which provides access to Tor Onion Services without using the Tor Browser.

The websites abuse.ch~\cite{abusech}, broadanalysis~\cite{broadanalysis} and malware-traffic-analysis~\cite{malwaretrafficanalysis} (referred to as data sources from here on) collect information about indicators of compromise (IOC) related to Cerber, such as domains or IP addresses. We download the domains used by Cerber on a daily basis and enrich them with information from the WHOIS database (obtained from a service called whoisxmlapi~\cite{whoisxmlapi}). We call the domains included in this dataset \textit{known Cerber domains} because they have been identified as Cerber domains in the past.

\begin{table}[tbh]
\centering
\caption{Examples of Cerber domains and  detection dates}
\label{cerber_domains_detection_dates}
\begin{tabular}{@{}lr@{}}
\toprule
\textbf{Domain}             & \textbf{Detection date} \\ \midrule
hjhqmbxyinislkkt.17rm9b.top & 2017-05-20     \\
hjhqmbxyinislkkt.1bas8q.top & 2017-03-02     \\
p27dokhpz2n7nvgr.1cbcpy.top & 2017-02-02     \\
avsxrcoq2q5fgrw2.1nsnuh.top & 2016-12-21     \\ \bottomrule
\end{tabular}
\end{table}

Table~\ref{cerber_domains_detection_dates} shows four examples of known Cerber domains. A typical Cerber domain consists of a 16 character long third-level domain, followed by a six character long second-level domain, followed by a top-level domain:

\vspace{1em}
$\underbrace{p27dokhpz2n7nvgr}_{third-level\ domain}.\underbrace{1cbcpy}_{second-level\ domain}.\underbrace{top}_{top-level\ domain}$
\vspace{1em}

The last time a known Cerber domain did not follow this schema was on December 5, 2016. Instead of a six character long second-level domain, Cerber used an .onion.to address. The usage of this domain provides access to the Tor2web service. However, this was the first time a divergent name schema was used since October 2, 2016 (eight characters long second level domain), so  it is reasonable to assume the above schema is currently default for Cerber domains.

\subsection{Dataset}
\label{sec:dataset}
\begin{figure*}[ht]
  %\centering
  \includegraphics[width=\textwidth]{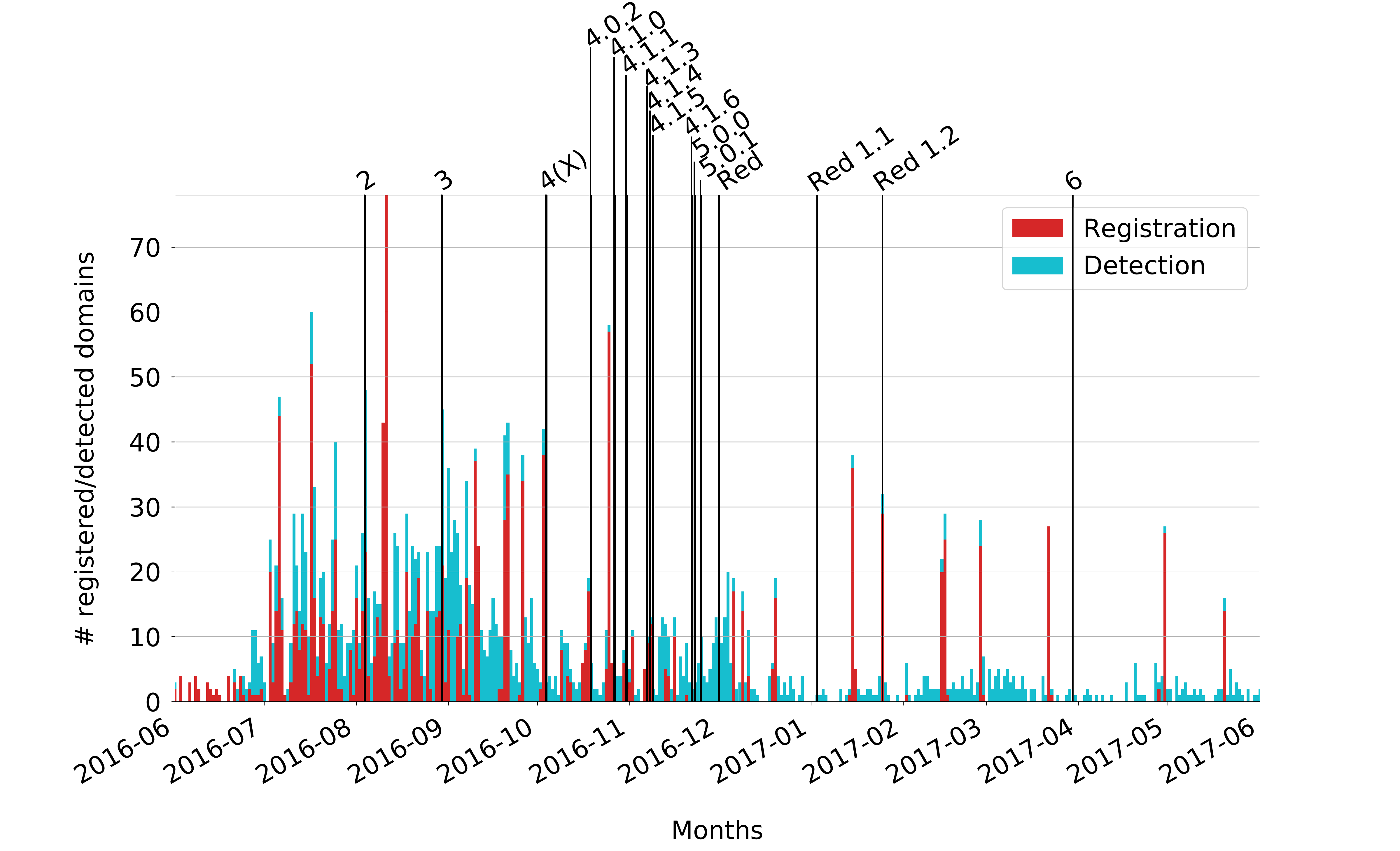}
  \caption{Number of detected and registered known Cerber domains per day}
  \label{cerber_detection_registration}
\end{figure*}

As of May 31, 2017, we collected 1,459 known Cerber domains from our data sources abuse.ch, broadanalysis and malware-traffic-analysis. The bar chart in Figure~\ref{cerber_detection_registration} shows the daily number of registered and detected known Cerber domains, where ``reg'' (registered) refers to the domain registration date obtained from WHOIS information, and ``det'' (detected)  means the number of  domains published by our data sources on that particular day. Additionally, the figure shows the release dates of new Cerber versions.

The number of registered and detected domains is not evenly distributed. Especially after the release of version 2 (denoted as 2)~\cite{bleeping_16_08_04} and version 3 (denoted as 3)~\cite{bleeping_16_08_31}, the number of detected domains increased considerably. Later on, Cerber released multiple major and minor versions in a short period of time so that it is challenging to link detected domains to a certain Cerber version. In 2017, Cerber released two minor versions Red 1.1~\cite{sensortech_17_01_03} and Red 1.2~\cite{sensortech_17_01_25} (Cerber stopped naming versions so that 1.1 and 1.2 are no official names). Later in 2017, Cerber developers released version 6~\cite{bleeping_17_05_03}. It is notable that especially in 2017, a high number of domains is registered only on a few days.

%\begin{figure*}[ht]
%  %\centering
%  \includegraphics[width=\textwidth]{fig/cerber_detection_time}
%  \caption{Number of detected known Cerber domains per day}
%  \label{cerber_detection_time}
%\end{figure*}

%The bar chart in Figure~\ref{cerber_registration_time} shows the daily number of registered  Cerber domains, based on WHOIS information, and the number of registered Cerber domain candidates per day. As of February 15th 2017, we found 412 Cerber domain candidates. The bar chart shows that a high number of both known Cerber domains and Cerber domain candidates are registered only on a few days. In advance of version 2 and version 3 a considerably higher number of domains was registered.

%\begin{figure*}[ht]
%  %\centering
%  \includegraphics[width=\textwidth]{fig/cerber_registration_time}
%  \caption{Number of registered known Cerber domains per day}
%  \label{cerber_registration_time}
%\end{figure*}

\begin{figure*}[ht]
  %\centering
  \includegraphics[width=\textwidth]{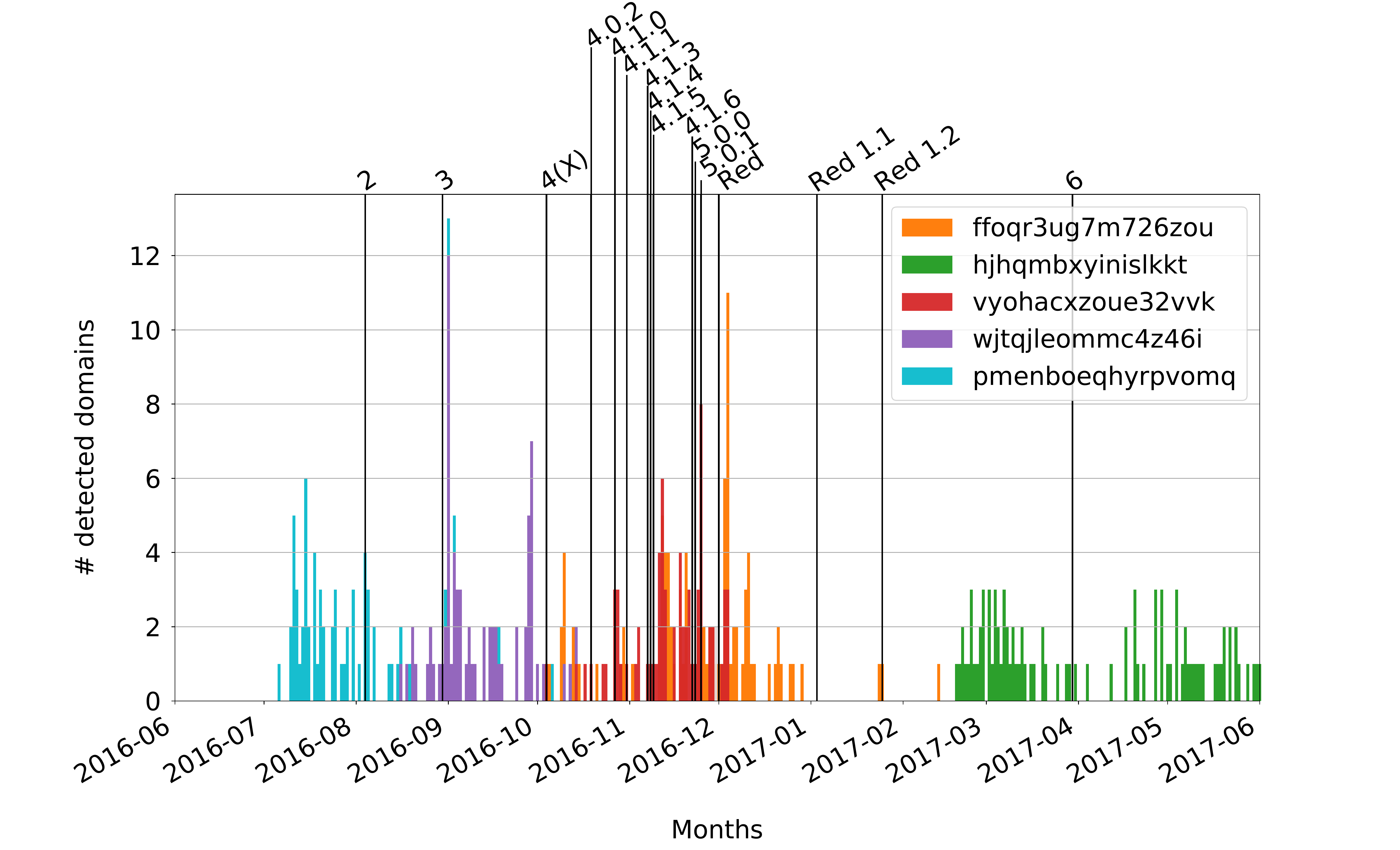}
  \caption{Number of detected known Cerber domains which were used with the particular third-level domain per day}
  \label{cerber_third_level_domain_detection_time}
\end{figure*}

Additionally, we analyzed Cerber's usage of third-level domains. It turned out that the 1459 known Cerber domains use only 20 different third-level domains. Figure~\ref{cerber_third_level_domain_detection_time} shows a selection of five of Cerber's third-level domains. Cerber uses each third-level domain only for a limited period of time. At first glance, the third-level domain might look like a good indicator to distinguish different campaigns or versions. However, Cerber uses for example ``p27dokhpz2n7nvgr'', a recent third-level domain, in a variety of different Cerber infections, i.e., via malicious spam emails~\cite{malspam-example} or exploit kits~\cite{exploit-kit-example} from different campaigns. Further investigations showed that each third-level domain belongs to a hidden service in the Tor network. 

Our analysis of the known Cerber domains shows that Cerber uses 32 different top-level domains. However, only the two different top-level domains \textit{bid} and \textit{top} appear in significant numbers (bid: 551 domains, top: 741 domains). Additionally, some domains were added multiple times to our data sources, e.g., with different third-level domains. Therefore, the distinct number of domains differs: 459 bid and 621 top domains. Table~\ref{cerber_prevalent_tlds} shows the number of \textit{top} and \textit{bid} domains detected per month between June 2016 and May 2017 (including double additions). According to our data sources, Cerber used the top-level domain \textit{bid} mostly between August 2016 and October 2016. By the beginning of 2017, Cerber stopped using domains of the top-level domain \textit{bid}. In contrast, Cerber used the top-level domain \textit{top} predominantly in two periods between July 2016 and August 2016 as well as since November 2016 to this day. Therefore, we focus our further analysis only on the top-level domain \textit{top}.

\begin{table*}
\centering
\caption{Number of detections per month for top-level domains \textit{bid} and \textit{top}}
\label{cerber_prevalent_tlds}
\begin{tabular}{lrrrrrrrrr}
\hline
\textbf{} & \textbf{2016-06} & \textbf{2016-07} & \textbf{2016-08} & \textbf{2016-09} & \textbf{2016-10} & \textbf{2016-11} & \textbf{2016-12} & \textbf{2017-01 -- 05} \\ \hline
bid       & 0                & 2                & 157              & 268               & 78               & 33               & 13               & 0                \\
top       & 9                & 187              & 101               & 3                 & 12               & 97               & 97               & 236              \\ \hline
\end{tabular}
\end{table*}

\subsection{Predicting Malicious Domains}
\label{sec:pred-mal-dom}
We evaluate RAPTOR within a real-world study that uses data from our data sources abuse.ch, malware-traffic-analysis and broadanalysis as training data, and zone file differences of the top-level domain \textit{top} we collected over the last couple of months as test data.
Specifically, we collected zone files from 2016/08/30, 2016/11/28, 2017/01/10 and 2017/01/16 until 2017/05/31 (due to technical issues, we missed the zone files from 2017/01/17, 2017/02/18 and 2017/05/14). However, the approach can handle missing zone files very well because the only difference is a higher number of newly registered domains in a difference between two non-consecutive zone files, i.e., a higher number of test data. Only when the time period between two zone files is very long, it is possible that in the meantime, domains were registered and already detected by our data sources. In such a case, our approach would predict the already detected domains because the approach is not aware that domains were already detected. We encountered  this problem with the differences between 2016/08/30 and 2016/11/28 as well as 2016/11/28 and 2017/01/10, respectively. Nevertheless, we decided to incorporate those zone files in our evaluation because they reveal insights into an earlier phase of Cerber's activity and further show the usefulness of our approach.

For each zone file difference, we identify the domains our data sources already published as known Cerber domains at the second date of the zone file difference. We use these domains, along with a similar number of benign, i.e., non-Cerber, domains %as training data
to train the classifiers. Afterwards, we use the classifiers to filter the zone file difference and predict which domains will be used by Cerber in the future. From today's point of view, we can search for the predicted domains %in the later on by our data sources detected domains to evaluate our prediction.
in our data sources at a later date to evaluate the predictions.

Table~\ref{real-world-overall-results} describes the overall results for the real-world study. We analyzed 132 zone file differences and reduced the total number of domains to 13,909 with RAPTOR's \textit{Step 1 Classifier} and to 2,126 Cerber domain candidates with the \textit{Step 2 Classifier}. From 2016/08/30 until today, our data sources collected 459 distinct known Cerber domains in the top-level domain \textit{top}. The domains predicted by RAPTOR's \textit{Step 1 Classifier} contain 386 of those domains (recall of 0.84). We target the low precision of 0.02 with RAPTOR's \textit{Step 2 Classifier} which reduces the number of predicted domains considerably, increases the precision to 0.17 and lowers the recall only negligible (0.82 compared to 0.84). 

\begin{table}[!t]
\centering
\caption{Overall results of real world study}
\label{real-world-overall-results}
\begin{tabular}{@{}ll@{}}
\toprule
Category                                       & Amount     \\ \midrule
\# analyzed zone file differences              & 132        \\
\# newly registered domains                    & 12156927 * \\
\# domains after \textit{Step 1 Classifier}    & 13909      \\
\# verified domains \textit{Step 1 Classifier} & 386        \\
\# domains after \textit{Step 2 Classifier}    & 2126       \\
\# verified domains \textit{Step 2 Classifier} & 378        \\ \bottomrule
\end{tabular}
\caption*{* Three zone file differences contain a huge amount of domains which were only observed in a single zone file. Aditionally, WHOIS requests to a selection of those domains revealed that the domains are not registered. Therefore, it is feasible to assume that the registration of those domains was later canceled.}
\end{table}

There are multiple reasons to explain the rather low precision values.
First, we evaluate our prediction only against the Cerber domains detected by our data sources without any guarantee that those data sources are comprehensive and capture every used Cerber domain. Therefore, we developed a system to verify that even more domains than detected by our data sources are in fact used by Cerber. We periodically sent HTTP requests to the Cerber domain candidates and checked both the availability and if the response contained indicators of Cerber. For that purpose, we use the following URL and replace \{cerber\_candidate\} with the Cerber domain candidates:

\vspace{1em}
$\underbrace{p27dokhpz2n7nvgr.\{cerber\_candidate\}.top}_{domain}/\underbrace{46...D2}_{id}$
\vspace{1em}

The first part is a recently used third-level domain. However, we detected that, with one exception, every other third-level domain works as well because Cerber forwards those requests to the recently used third-level domain. The last part is a personal ID of a victim we found with a Google search. This ID is necessary to access the payment website. If we receive a successful response, we can check if the title tag contains the word ''Cerber`` and, thus, make sure it is actually one of Cerber's payment websites. Overall, we could verify 126 domains between February 2, 2017 and May 31, 2017. 99 of those domains were later added to the data sources so that we could verify another 27 domains to be used by Cerber.

\begin{figure*}[ht]
  \centering
  \includegraphics[width=0.9\textwidth]{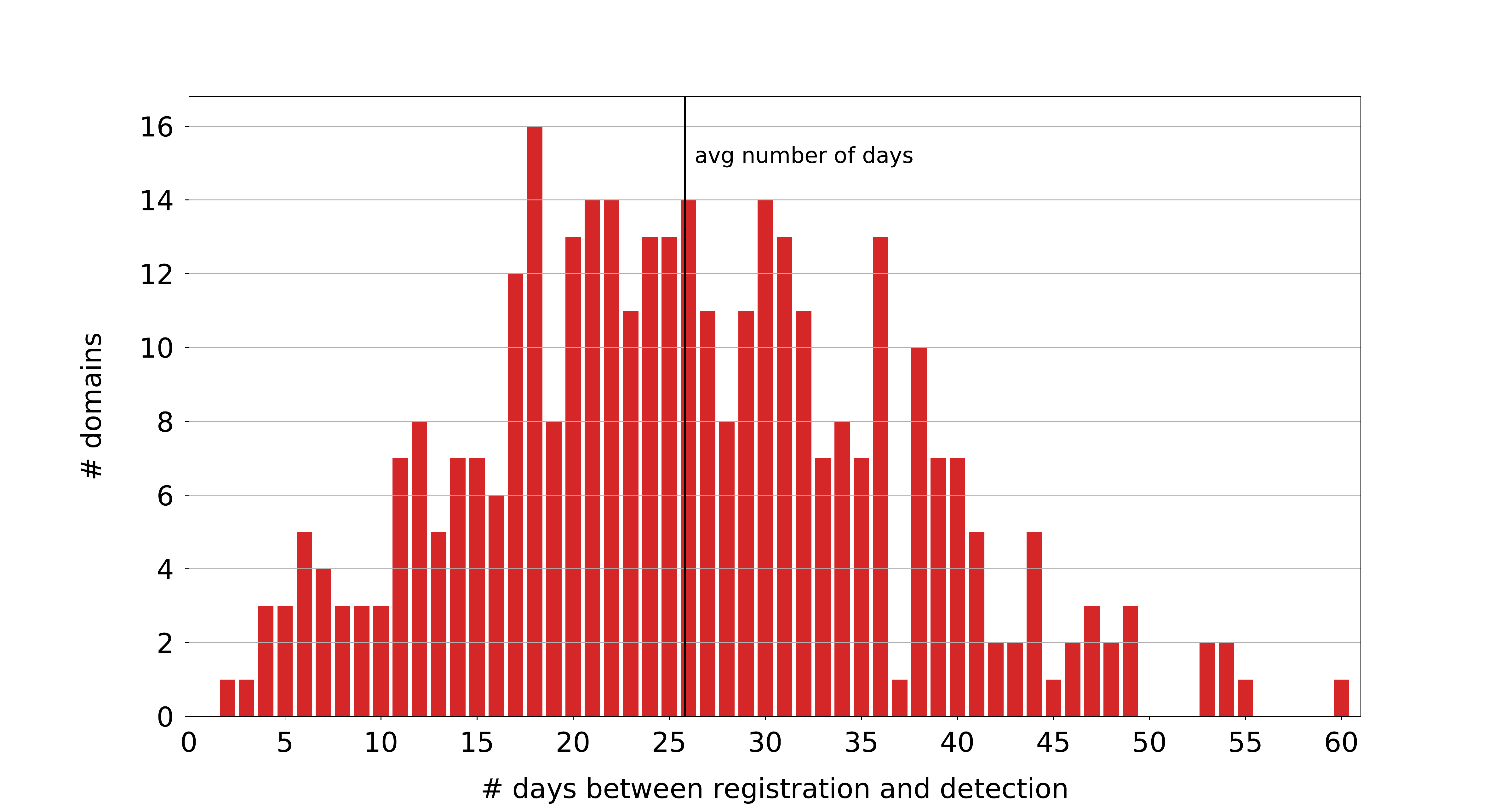}
  \caption{Number of domains and how many days after registration they were added to our datasources}
  \label{days_between_reg_det}
\end{figure*}

Additionally, another reason explaining the low precision is the number of days between registration of a domain and addition to our data sources. Figure~\ref{days_between_reg_det} shows our evaluation of the number of days between registration of a domain and addition to our data sources. It takes up to 60 days and on average 26 days pass by between registration and addition. Therefore, it is likely that additional domains we predicted will be added to our data sources in the future.

Furthermore, ransomware developers register a large number of domains at once without necessarily using all of them. Thus, we might predict domains actually being registered by Cerber developers, but cannot verify them because they are not actively used yet by the attackers, which in turn lowers our precision.

\subsubsection{Cross-Validation}
Additionally, we perform a cross-vali\-da\-tion to address the issues of the real-world study and show the feasibility of our approach. We use randomly chosen 10\% of the known Cerber domains as well as 10\% of the benign domains as test data and the remaining 90\% of both known Cerber domains and benign domains as training data. In contrast to our real-world evaluation, the cross-validation features the advantage that the test data is labeled, i.e., we know whether a domain is a known Cerber domain or not so that we can immediately verify the prediction results of our approach. We perform the cross-validation 100 times and calculate precision, recall and f1-score for each run. Table~\ref{cross-validation-results} shows minimum, maximum and average values for precision, recall and f1-score.

The high precision shows that our approach has a very low false positive rate, which is especially desired because the vast majority of newly registered domains is not used by Cerber or ransomware in general. Therefore, the approach works well to filter out the irrelevant newly registered domains. Additionally, the high recall reveals a good ability to find Cerber domains. Furthermore, it shows that the low recall in the  real-world evaluation is biased by the high number of not verifiable domains.

\begin{table}[!t]
\centering
\caption{Precision, recall and f1-score for cross-validation}
\label{cross-validation-results}
\begin{tabular}{llll}
\hline
          & \textbf{min.} & \textbf{max.} & \textbf{avg.} \\ \hline
precision & 0.90 & 1.00 & 0.96 \\
recall    & 0.66 & 0.84 & 0.75 \\
f1-score  & 0.79 & 0.90 & 0.85 \\ \hline
\end{tabular}
\end{table}

\subsection{Predicting Detected Domains}
We use the time series forecasting models HMM, ARIMA, and ARIMAX (see Sec.~\ref{sec:ts-models}), to predict the daily number of domains involved in Cerber attacks that will be detected by our data sources (i.e., Cerber attacks). We collect and curate the time series of domains involved in Cerber attacks from the data sources described in Sec.~\ref{sec:data-source} for the period between August 30, 2016 and May 31, 2017. We leverage the output (time series of domain registrations) of our malicious domain prediction method (see Sec.~\ref{sec:pred-mal-dom}) as an external signal. We align the external signal with the Cerber attacks using correlation analysis~\cite{rabiner1975theory}. 
%The baserate, HMM, ARIMA use historical  Note that ARIMAX is the only model that can take advantage of external signals. 
ARIMAX exploits both historical time series of domains involved in Cerber attacks and the external signals, whereas the baserate, HMM, and ARIMA only use the time series of domains involved in Cerber attacks for prediction. 

%The time series approach can be easily generalized to other types of ransomware, and indeed, to other prediction tasks, such as predicting the number of domains involved in ransomware attacks, not just the number of malicious domain registrations. We illustrate the flexibility of the approach on the task of forecasting the number of attacks reported by ransomware tracker~\cite{abusech}. These attacks can be grouped into nine malware types (see Table~\ref{tab:ransom-types}). 
%Besides Cerber, we also use Locky in the prediction task, as these types contain a large enough number of events for learning model parameters. 
%Similar to Sec.~\ref{sec:results-dom-reg}, 

%First, we use only the detected domains in our data sources in the prediction task. 

\begin{figure*}[!t]
  \centering
  \includegraphics[scale=0.38]{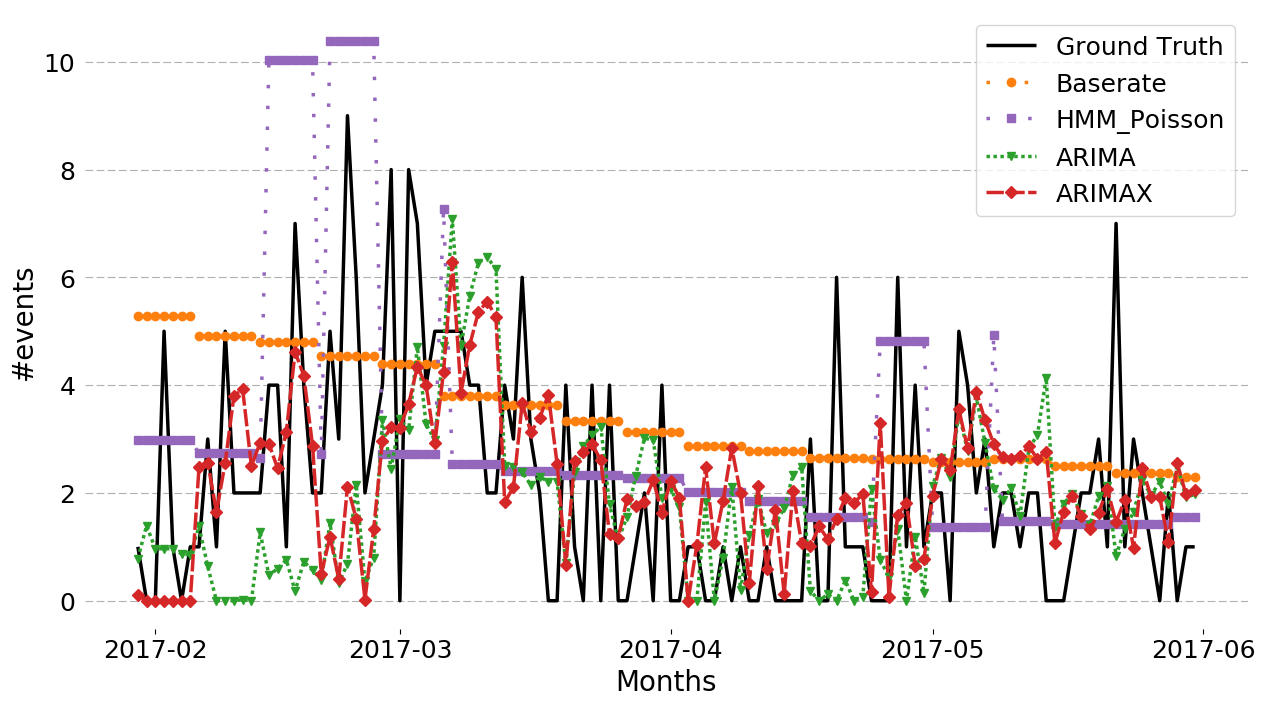}
  \caption{Forecasting the number of detected Cerber ransomware domains using HMM, ARIMA, ARIMAX, and a base rate model. This figure illustrates the predicted values over the test set (last 40\% of the time series data) using sliding window approach with look-ahead of 7 days. Here HMM with two hidden states and Poisson emission probability is used, and optimal ARIMA and ARIMAX models are identified using grid search over parameters.}
  \label{fig:cerber-attacks}
\end{figure*}

For prediction, we divide the time series of domains involved in attacks in the data sources into training (first 60\% of data) and test (last 40\% of data) sets. We learn the models' parameters on the training set and forecast the expected number of detected domains involved in Cerber attacks for the next seven days (in the test set). We then shift the training window to the right by seven days and repeat the procedure. We pick the training window, number of hidden states, emission probability distribution, and orders of autoregressive models using a grid search for best scores. 
Fig.~\ref{fig:cerber-attacks} shows the  expected number of domains involved in attacks predicted with the three proposed methods and a base rate model. For Fig.~\ref{fig:cerber-attacks}, we use an HMM with two hidden states and Poisson emission probability, and for ARIMA and ARIMAX, we apply a grid search over $(max_p=7, max_d=2, max_q=7)$ to identify the optimal model in terms of AIC score. Table~\ref{tab:cerber-attack} 
presents corresponding quantitative comparisons between three methods. ARIMAX outperforms other models for Cerber type ransomware attacks. This results demonstrates that exploitation of external signals helped ARIMAX better forecast when there is not historical data for attacks. 

%Interestingly, predicting that the number of domains involved in future attacks will be the same as today performs fairly well (Nash Sutcliffe score 0.38), in contrast to domain registrations (Nash Sutcliffe score -0.46). However, the proposed models handily beat this simple base rate model for both ransomware types.

% \begin{figure*}[!t]
%   \centering
%   %\includegraphics[scale=0.47]{fig/cerber_pred_simple_v2_sm_w_50_p_3_d_1_q_0}
%   \includegraphics[scale=0.47]{fig/cerber_attack_pred_cfg_02.png}
%   \caption{Forecasting the expected number of Cerber ransomware attacks using HMM, ARIMA, and a base rate model. This figure illustrates the predicted values over the test set. Here HMM with three hidden states and Poisson emission probability are used, and optimal ARIMA model is identified using grid search over parameters}
%   \label{fig:cerber-attacks}
% \end{figure*}

% \begin{table}[!t]
% \centering
% \caption{Forecasting of Cerber ransomware attacks using hidden Markov model (HMM), autoregressive model (ARIMA). Methods are compared in terms of different performance metrics: mean absolute error (MAE), root mean squared error (RMSE), and mean absolute scaled error (MASE).}
% \label{tab:cerber-attack}
% \begin{tabular}{@{}lrrrr@{}}
% \toprule
% %\textbf{Measures} & \textbf{Current Count} & \textbf{Rolling Avg} & \textbf{HMM} & \textbf{ARIMA} \\ \midrule
% \textbf{Measures} & \textbf{Baserate} & \textbf{HMM} & \textbf{ARIMA} \\ \midrule
% %\multicolumn{5}{c}{(b) Cerber}\\ \midrule
% %MSE            & 21.20          & 31.89       & 20.19         & \textbf{19.63} \\
% %NMSE           & 0.52          & 0.71        & 0.49          & \textbf{0.48}  \\
% MAE            & 2.71         & 2.10           & \textbf{1.63}  \\
% RMSE           & 3.19         & 2.62          & \textbf{2.23}  \\
% %NRMSE          & \textbf{0.16} & 0.20         & \textbf{0.16} & \textbf{0.16}  \\
% %Scatter Index  & 0.73          & 0.89        & 0.71          & \textbf{0.70}   \\
% %MAE            & 3.36          & 4.54        & 3.50           & \textbf{3.28}  \\
% MASE           & 1.61         & 1.24          & \textbf{0.97}  \\
% %MEDAE          & 3.00           & 4.05        & 2.82          & \textbf{2.59}  \\
% %Nash Sutcliffe & 0.38          & 0.07        & 0.41          & \textbf{0.42}  \\ 
% \bottomrule
% \end{tabular}
% \end{table}

\begin{table}[!t]
\centering
\caption{Forecasting of Cerber ransomware attacks using hidden Markov model (HMM) and autoregressive models (ARIMA and ARIMAX). Methods are compared in terms of different performance metrics: mean absolute error (MAE), root mean squared error (RMSE), and mean absolute scaled error (MASE).}
\label{tab:cerber-attack}
\begin{tabular}{@{}lcccc@{}}%
\toprule%
\textbf{Measure}&\textbf{Baserate}&\textbf{HMM\_Poisson}&\textbf{ARIMA}&\textbf{ARIMAX}\\%
\midrule%
MAE&2.05&2.15&1.81&\textbf{1.66}\\%
RMSE&2.41&2.83&2.37&\textbf{2.10}\\%
MASE&1.11&1.16&0.98&\textbf{0.90}\\%
\bottomrule%
\end{tabular}%
\end{table}

\section{Discussion and Conclusions}
\label{conclusion}
The rising popularity of ransomware has challenged the ability of antivirus companies to keep up with the proliferation of malware variants. This was dramatically demonstrated on May 17, 2017, when a vulnerability in Microsoft Windows operating systems enabled a massive attack by the WannaCry ransomware that crippled infrastructure and emergency services in many countries around the world. The growing ransomware threat calls for new defensive solutions. 

In this paper, we presented RAPTOR, a promising approach that relies on \emph{predicting} properties of ransomware attacks. 
Our approach leverages an intuition that attackers facilitate their attacks through some processes. By fingerprinting these processes, we can model the behavior of the attackers and use the models to predict future attacks. To paraphrase an old saying, ``foreknowledge is the best defense''.
% KL added significance of prediction
Anticipating properties of future ransomware attacks, or even simply their number, can help enterprises and individuals to better manage their resources defending against attacks. For example, knowing there will be a spike in the number of ransomware attacks in the coming days can lead enterprises to deploy additional spam detection services and warn their users to be vigilant.

Our approach is based on the life-cycle of an attack. First, attackers need to infect the victim's system; then, they need the compromised system to communicate with their payment infrastructure to receive the ransom. To remain clandestine, this requires attackers to continuously register new domains. In a first step, RAPTOR learns features common to malicious domains by looking at structure and WHOIS information of example domains involved in ransomware attacks. Then, by monitoring newly registered domains, we can flag the ones that exhibit the incriminated features. 

We built upon the intuition that the attackers' actions are not random and independent, but temporally correlated. Thus, in a second step, RAPTOR uses time series prediction methods to learn a model of an attacker based on historical events, and use the model to predict when domains are involved in attacks. We described several algorithms used in time series analysis---HMM and ARIMA---which we applied to model the prediction of the number of domains involved in detected attacks. These algorithms take as input historical sequences of events (registrations or detections) and predict the likely number of new events. Moreover, we described another algorithm---ARIMAX---that uses predicted malicious domains as an external signal in time series prediction in addition to historical sequences of events. 

We evaluated RAPTOR by applying it to predict the stages of Cerber, a popular ransomware. By applying our feature extraction approach to zone files  of the top level domain $top$ over a period starting from August 30, 2016 to May 31, 2017, we predicted 2,126 Cerber domain candidates. Of those domains, 378 were later confirmed to be malicious, due to their  appearance in our data sources \textit{abuse.ch}, \textit{broadanalysis} and \textit{malware-traffic-analysis}. 
In addition, by sending requests to the domains on a periodic basis, we additionally verified that Cerber used 126 of these domain candidates. However, only 99 of them were later published by our data sources and are already included in the 378 domains mentioned above. Hence, 27 of the domains used by Cerber were not detected by any of our data sources, which supports our assumption that Cerber uses more domains than detected by our data sources.
% Perhaps a stronger statement here
The remaining %402 
registered  but not verified domains have not been used yet. Nonetheless, Cerber starts using earlier-registered domains on a daily basis, thus the number of verified domains is likely to increase in the future. Additionally, the cross-validation showed that RAPTOR has in contrast to the real-world evaluation a very low false positive rate.

%In our evaluation, we focused only on the ransomware Cerber. However, RAPTOR works with other ransomware families as well if a set of example domains is available to train the classifiers and the domains used by the ransomware share common characteristics, such as domain length or patterns in registration information. Instead of always using the same features, it might be useful to analyze the example domains further to estimate whether a certain feature is useful or not, e.g., the domain's length is not a good feature if the example domains' lengths vary a lot. Furthermore, additional features might be feasible depending on the example domains. Unfortunately, both estimating the usefulness of existing and the discovery of new features is difficult to do in an automated way so that some manual effort is necessary to use RAPTOR with other ransomware families.

Additionally, characteristics of ransomware domains pose challenges for time series prediction. As can be seen in Fig.~\ref{fig:cerber-attacks}, Cerber ransomware activity changed dramatically at the beginning of 2017. This non-stationarity makes leveraging old patterns difficult. This kind of volatility is a possible reason for HMM not performing as well as other methods.

While promising, these approaches have limitations.
The feature extraction based approach showed very good precision. However, an adversary ransomware developer could avoid detection by using, at the time of domain registration, different domain name schemes and WHOIS information that is hard to cross-reference. However, it is difficult to register a high number of domains in an automated way without having any detectable patterns.
Time series prediction, while being flexible and generalizable to new domains, also has limitations. Data sparsity remains a challenge for time series prediction methods: when there is not enough signal, results are not reliable. However, despite these shortcomings, forecasting cyber threats remains a promising new tool for cyber defense. 
\section{Acknowledgments}
%Anonymized for review
This project was funded by the Office of the Director of National Intelligence (ODNI) and the Intelligence Advanced Research Projects Activity (IARPA) via the Air Force Research Laboratory (AFRL) contract number FA8750-16-C-0112. The U.S. Government is authorized to reproduce and distribute reprints for Governmental purposes notwithstanding any copyright annotation thereon.
Disclaimer: The views and conclusions contained herein are those of the authors and should not be interpreted as necessarily representing the official policies or endorsements, either expressed or implied, of ODNI, IARPA, AFRL, or the U.S.
Government.

\bibliographystyle{splncs}
\bibliography{./bibliography}

\end{document}